\documentclass[aps,prl,twocolumn,superscriptaddress,showpacs,preprintnumbers,amsmath,amssymb]{revtex4}

\usepackage{graphicx} 
\usepackage{dcolumn}  
\usepackage{color}  
\usepackage{multirow}

\begin{document}



\title{ \quad\\[1.0cm]
Evidence for {\boldmath $B^- \to 
D^{+}_s K^- \ell^- \bar{\nu}_{\ell}$} and search for {\boldmath $B^- \to 
D^{*+}_s K^- \ell^- \bar{\nu}_{\ell}$} }

\affiliation{University of Bonn, Bonn}
\affiliation{Budker Institute of Nuclear Physics SB RAS and Novosibirsk State University, Novosibirsk 630090}
\affiliation{Faculty of Mathematics and Physics, Charles University, Prague}
\affiliation{University of Cincinnati, Cincinnati, Ohio 45221}
\affiliation{Department of Physics, Fu Jen Catholic University, Taipei}
\affiliation{Gifu University, Gifu}
\affiliation{Hanyang University, Seoul}
\affiliation{University of Hawaii, Honolulu, Hawaii 96822}
\affiliation{High Energy Accelerator Research Organization (KEK), Tsukuba}
\affiliation{Indian Institute of Technology Guwahati, Guwahati}
\affiliation{Indian Institute of Technology Madras, Madras}
\affiliation{Institute of High Energy Physics, Chinese Academy of Sciences, Beijing}
\affiliation{Institute of High Energy Physics, Vienna}
\affiliation{Institute of High Energy Physics, Protvino}
\affiliation{Institute for Theoretical and Experimental Physics, Moscow}
\affiliation{J. Stefan Institute, Ljubljana}
\affiliation{Kanagawa University, Yokohama}
\affiliation{Institut f\"ur Experimentelle Kernphysik, Karlsruher Institut f\"ur Technologie, Karlsruhe}
\affiliation{Korea Institute of Science and Technology Information, Daejeon}
\affiliation{Korea University, Seoul}
\affiliation{Kyungpook National University, Taegu}
\affiliation{\'Ecole Polytechnique F\'ed\'erale de Lausanne (EPFL), Lausanne}
\affiliation{Faculty of Mathematics and Physics, University of Ljubljana, Ljubljana}
\affiliation{Luther College, Decorah, Iowa 52101}
\affiliation{University of Maribor, Maribor}
\affiliation{Max-Planck-Institut f\"ur Physik, M\"unchen}
\affiliation{University of Melbourne, School of Physics, Victoria 3010}
\affiliation{Graduate School of Science, Nagoya University, Nagoya}
\affiliation{Kobayashi-Maskawa Institute, Nagoya University, Nagoya}
\affiliation{Nara Women's University, Nara}
\affiliation{National Central University, Chung-li}
\affiliation{Department of Physics, National Taiwan University, Taipei}
\affiliation{H. Niewodniczanski Institute of Nuclear Physics, Krakow}
\affiliation{Nippon Dental University, Niigata}
\affiliation{Niigata University, Niigata}
\affiliation{University of Nova Gorica, Nova Gorica}
\affiliation{Osaka City University, Osaka}
\affiliation{Pacific Northwest National Laboratory, Richland, Washington 99352}
\affiliation{Research Center for Electron Photon Science, Tohoku University, Sendai}
\affiliation{University of Science and Technology of China, Hefei}
\affiliation{Seoul National University, Seoul}
\affiliation{Sungkyunkwan University, Suwon}
\affiliation{School of Physics, University of Sydney, NSW 2006}
\affiliation{Tata Institute of Fundamental Research, Mumbai}
\affiliation{Excellence Cluster Universe, Technische Universit\"at M\"unchen, Garching}
\affiliation{Toho University, Funabashi}
\affiliation{Tohoku Gakuin University, Tagajo}
\affiliation{Tohoku University, Sendai}
\affiliation{Department of Physics, University of Tokyo, Tokyo}
\affiliation{Tokyo Institute of Technology, Tokyo}
\affiliation{Tokyo Metropolitan University, Tokyo}
\affiliation{Tokyo University of Agriculture and Technology, Tokyo}
\affiliation{CNP, Virginia Polytechnic Institute and State University, Blacksburg, Virginia 24061}
\affiliation{Wayne State University, Detroit, Michigan 48202}
\affiliation{Yamagata University, Yamagata}
\affiliation{Yonsei University, Seoul}

  \author{J.~Stypula}\affiliation{H. Niewodniczanski Institute of Nuclear Physics, Krakow} 
  \author{M.~Rozanska}\affiliation{H. Niewodniczanski Institute of Nuclear Physics, Krakow} 

  \author{I.~Adachi}\affiliation{High Energy Accelerator Research Organization (KEK), Tsukuba} 
  \author{K.~Adamczyk}\affiliation{H. Niewodniczanski Institute of Nuclear Physics, Krakow} 
  \author{H.~Aihara}\affiliation{Department of Physics, University of Tokyo, Tokyo} 
  \author{D.~M.~Asner}\affiliation{Pacific Northwest National Laboratory, Richland, Washington 99352} 
  \author{T.~Aushev}\affiliation{Institute for Theoretical and Experimental Physics, Moscow} 
  \author{A.~M.~Bakich}\affiliation{School of Physics, University of Sydney, NSW 2006} 
  \author{V.~Bhardwaj}\affiliation{Nara Women's University, Nara} 
  \author{B.~Bhuyan}\affiliation{Indian Institute of Technology Guwahati, Guwahati} 
% ...
\collaboration{The Belle Collaboration}


\begin{abstract}
We report measurements of the decays 
$B^-\to D^{(*)+}_s K^-\ell^- \bar{\nu}_{\ell}$ in a 
data sample 
containing $657\times10^6$ $B\bar{B}$ pairs 
collected with the Belle detector at the KEKB 
asymmetric-energy $e^+e^-$ 
collider. We {observe} a signal with a 
significance of 
6$\sigma$ for the combined $D_s$ and $D^{*}_s$ modes and
find the first evidence of the $B^-\to D_s^+K^-\ell^-\bar{\nu}_{\ell}$
decay with a significance of $3.4\sigma$. 
We measure the following branching fractions: 
$\mathcal{B}(B^-\to 
D^{+}_sK^-\ell^-\bar{\nu}_{\ell})
=(0.30\pm 0.09({\rm stat})^{+0.11}_{-0.08}({\rm syst}))\times 10^{-3}$ and $\mathcal{B}(B^-\to D^{(*)+}_sK^-\ell^-\bar{\nu}_{\ell})=
(0.59\pm 0.12({\rm stat})\pm 0.15({\rm syst}))\times 10^{-3}$
and set an upper limit 
$\mathcal{B}(B^-\to 
D^{*+}_sK^-\ell^-\bar{\nu}_{\ell})
<0.56 \times 10^{-3}$ at the $90\%$ confidence level.
We also present the first measurement of the $D_s^+K^-$ invariant mass 
distribution in these decays, which is dominated by a {prominent peak
around} $2.6$ GeV$/c^2$.
\end{abstract}

\pacs{13.20.He, 14.40.Nd}

\maketitle

\tighten

{\renewcommand{\thefootnote}{\fnsymbol{footnote}}}
\setcounter{footnote}{0}

Semileptonic $B$ decays play a key role
in testing the Standard Model 
(SM) and in the understanding of heavy quark dynamics.
In particular, they are used to determine the weak mixing parameters 
$|V_{qb}|$ $(q=c,u)$, complementing the measurements
of $CP$ asymmetries used to verify the Cabibbo-Kobayashi-Maskawa (CKM) mechanism of the SM \cite{CKM}.
The tension at the level of $2$ standard deviations ($\sigma$) between the values of 
$|V_{qb}|$ extracted from
inclusive and exclusive $B$ decays \cite{HFAG}, as well as some discrepancies between
measurements and theoretical expectations for semileptonic $B$ decays to
excited charmed mesons, 
may indicate problems 
in the theoretical tools
or in the interpretation of the experimental results.

Semileptonic $B$ decays to final states containing a ${D}_s^{(*)+}\bar K$ system {\cite{CC}}
 provide information about the poorly known 
region of hadronic masses above $2.46~{\rm GeV}/c^2$,
covering radially excited $D$ meson states \cite{Dst}.
Further exploration of this region may help solving some puzzles in semileptonic $B$ decays \cite{puzzles}.
Recently, BaBar reported
an observation of $B^-\to D^{(*)+}_s K^- 
\ell^-\bar{\nu}_{\ell}$ 
(which did not distinguish between the $D_s$ 
and $D_s^{*}$ final states)
with a branching 
fraction of $\mathcal{B}(B^-\to D^{(*)+}_s K^- 
\ell^-\bar{\nu}_{\ell})= (6.13^{+1.04}_{-1.03}({\rm stat})\pm 0.43({\rm syst})
\pm 0.51 (\mathcal{B}(D_s)))\times 10^{-4}$ \cite{BaBar}.

In this paper, we present measurements of
$B^-\to D^{+}_s K^- \ell^- \bar{\nu}_{\ell}$ and
$B^-\to D^{*+}_s K^- \ell^- \bar{\nu}_{\ell}$
decays using a data sample 
containing $657\times 10^6$ $B\bar{B}$ pairs that 
were collected with 
the Belle detector at the KEKB asymmetric-energy $e^+e^-$
collider \cite{KEKB} operating at the $\Upsilon(4S)$ resonance 
(center-of-mass energy $\sqrt{s}=10.58$ GeV). 
The Belle 
detector is a large-solid-angle magnetic spectrometer consisting of a 
silicon vertex detector, a 50-layer central drift chamber, a 
system of aerogel Cherenkov counters, time-of-flight scintillation 
counters and an electromagnetic calorimeter comprised of CsI(Tl) 
crystals located inside a superconducting solenoid coil that 
provides a 1.5~T magnetic field. An iron {flux return} located outside the 
coil is instrumented to identify $K_L^0$ mesons 
and muons.
A detailed description 
of the detector can be found in Ref.~\cite{Belle}. 
We use Monte Carlo (MC) simulations to estimate signal
efficiencies and background contributions.
Large {signal samples of} 
$B^-\to D^{(*)+}_s K^-\ell ^- \bar{\nu}_{\ell}$ decays are generated 
with 
the 
EvtGen package \cite{evtgen},
using a phase space model and the ISGW2 model \cite{isgw2} including the 
resonances that can decay to $D^{(*)}_s\bar{K}$. 
Radiative effects are modeled by PHOTOS \cite{photos}. 
MC samples equivalent to  
about ten (six) times the accumulated data are 
used to evaluate the background from 
$B\bar{B}$ (continuum $q\bar{q}$, where $q=u,d,s,c$) events.  

In the analysis, we use charged tracks with impact parameters that are 
consistent with an origin at the beam spot and have transverse momenta 
above 50 MeV/$c$.
Masses are assigned
using information from particle 
identification subsystems.  
The efficiency for kaon (pion) identification ranges from $84\%$ to 
$98\%$ ($92\%$ to $94\%$) depending on the track momentum with a pion 
(kaon) misidentification probability of about $8\%$ 
($16\%$). 
Electrons and muons are selected with an efficiency of about 90\% and 
a misidentification rate below 0.2\% ($e$) and 
1.4\% ($\mu$). 
The momenta of particles identified as electrons  are 
corrected for bremsstrahlung by adding photons within a 50 mrad cone
around the charged particle's trajectory.

$D_s^+$ candidates are reconstructed in the cleanest decay chain: 
$D^+_s\to \phi \pi^+$, $\phi\to K^+K^-$ {($2.32 \pm 0.14 \%$ product branching fraction)} and subjected to a vertex fit.
We accept candidates in the invariant mass range
of $1.934$ GeV/$c^2 < M_{D_s} <2.003$ GeV/$c^2$,
and define the signal window
within $\pm 14$ ${\rm MeV}/c^2$ around 
the world average $D_s$ mass \cite{PDG}.
The width of this window corresponds to $4\sigma$ {of the reconstructed $D_s$ mass},
using the resolution determined from control samples in data (mentioned later).
The regions outside the signal window are considered as $M_{D_s}$ sidebands.
$D_s^+$ candidates are combined with
photons with an energy $E_{\gamma}>125~{\rm MeV}$ 
to form $D_s^{*+}$ candidates, subjected to a mass constrained vertex fit.
Throughout this paper, all kinematic variables are defined in 
the $\Upsilon(4S)$ rest frame,
unless otherwise stated.
$D_s^{*+}$ candidates with an invariant mass in the
range of $2.079$ GeV/$c^2 < M_{D_s^{*}} < 2.155$ ${\rm GeV}/c^2$ are 
accepted for further analysis.
The signal window is defined as  $2.087$ GeV/$c^2 < M_{D_s^{*}} < 2.137$ ${\rm 
GeV}/c^2$ ($3.7\sigma$ in $M_{D_s^{*}}$).
Signal candidates for the decays considered here ($B_{\rm sig}$) are formed by combining
a negatively charged kaon and lepton ($e$ or $\mu$) 
with a $D^{(*)+}_s$ candidate.
In the case of multiple 
$B_{\rm sig}$ candidates {(22\% of events after final selection requirements have multiple $B_{\rm sig}$ candidates)}, the one 
with the greatest confidence level of the vertex fit is chosen. 
Events with accepted $D^{*+}_s K^-\ell^-$ candidates ($D^{*}_s$ 
sample) are removed
from the set of $D_s^+K^-\ell^-$ candidates ($D_s$ sample).
Another charge configuration,
$D^{(*)+}_s K^+ \ell^-$,
populated
by decays of the type $B \to D_s^{(*)+} \bar D^{(*)}$,
$\bar D \to \ell^- \bar \nu_{\ell} K^+ X$, 
is used as a 
control sample.

Signal events are identified using 
the variable $X_{\rm mis}$,
introduced in Ref. 
\cite{Matyja} {and} defined as:
$X_{\rm mis}\equiv (E_{\rm beam}-E_{D_s K \ell} 
-|\vec{p}_{D_s K \ell}|)/\sqrt{E_{\rm beam}^2-m^2_{B^+}}$,
where $E_{\rm beam}$ is the beam energy, 
$E_{D_s K \ell}$ and $\vec{p}_{D_s K \ell}$
denote the total energy and momentum of the $D_s K \ell$ system, respectively,
and $m_{B^+}$ is the nominal $B^+$ mass.
For {decays} with 
at most one 
massless invisible particle, as expected for the signal,
$X_{\rm mis}$ 
takes values in the range of $[-1,1]$, 
defined as the signal region,
while the background has a much broader 
distribution. 
$X_{\rm mis}$ is calculated with the 
four-momentum of the $D_s$ both in the $D_s$ 
and $D^{*}_s$ samples,
causing a small shift
of $X_{\rm mis}$ toward higher values for the $D^{*}_s$ 
case due to the {additional} low-energy photon.
With this definition, the $X_{\rm mis}$ distribution is more robust against imperfect modeling of photon spectra in MC and simplifies the signal extraction.

\begin{figure}
                \centering
                {\includegraphics[width=0.23\textwidth]{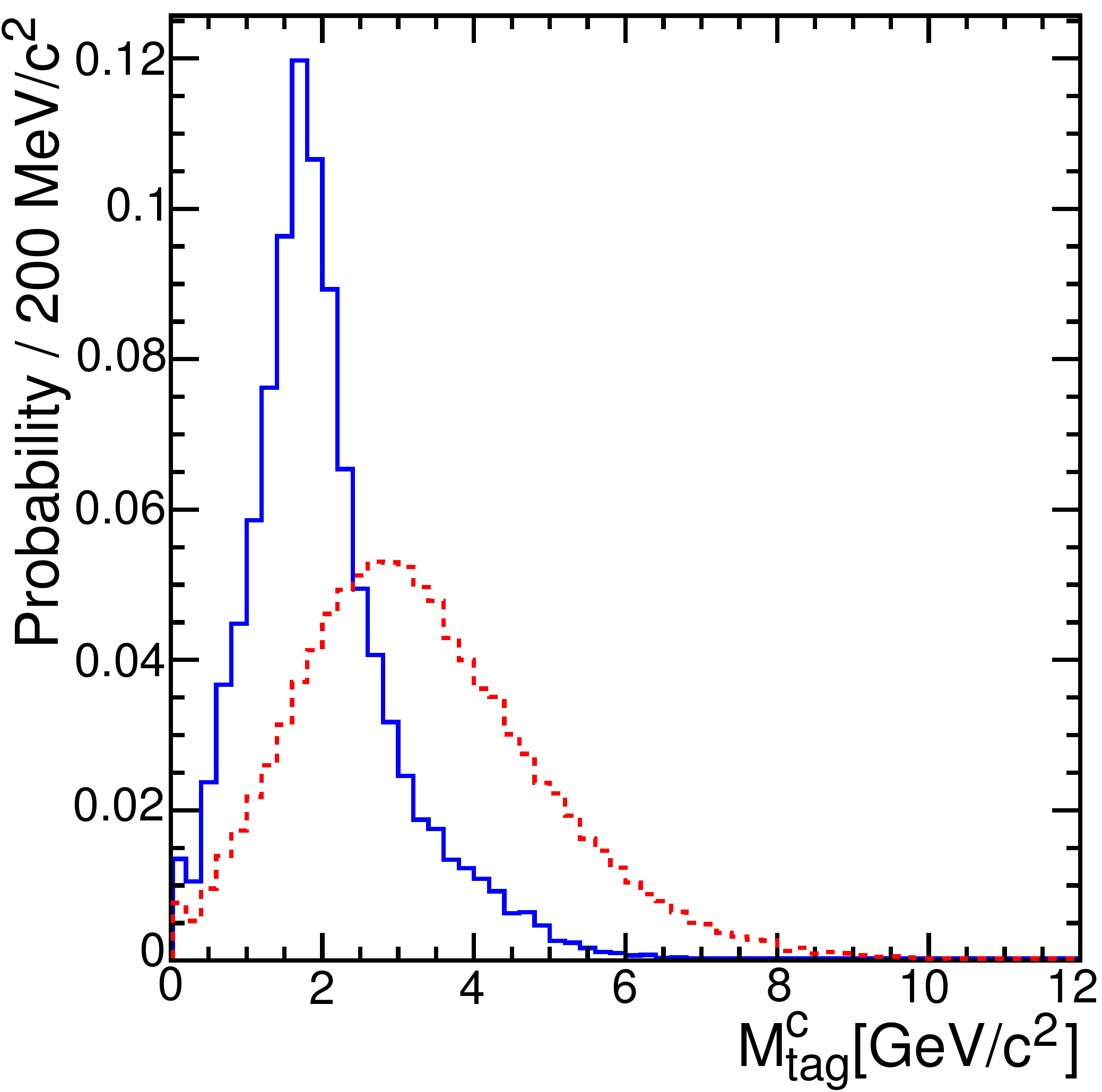}}  
                {\includegraphics[width=0.23\textwidth]{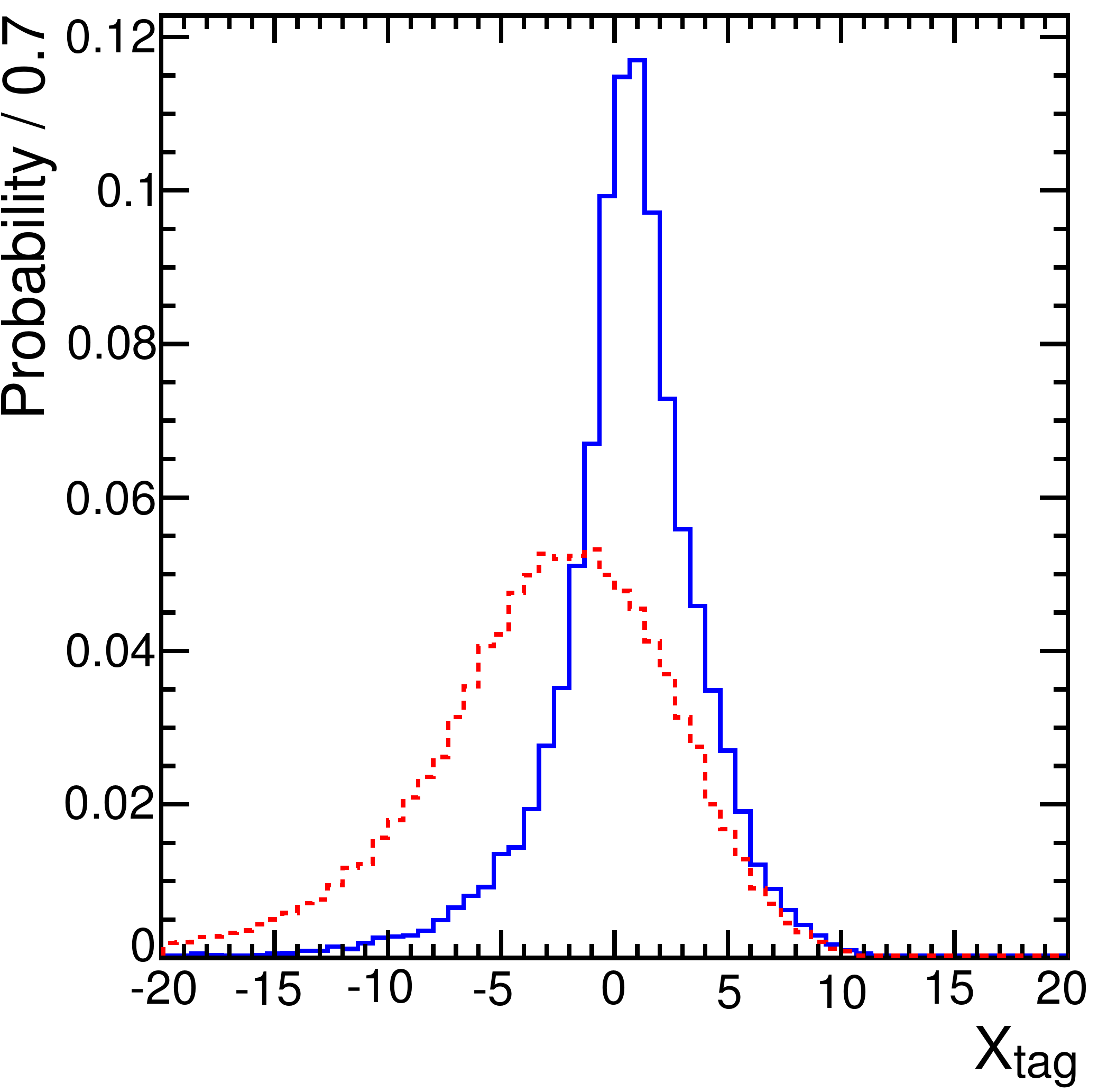}}
\caption{$M^c_{\rm tag}$ (left) and $X_{\rm tag}$ (right) distributions
for signal (blue, solid) and background (red, dashed) MC.}
\label{tcuts}
\end{figure}

Particles not assigned to the $B_{\rm sig}$ are used to 
reconstruct the tagging side of the event ($B_{\rm tag}$).
Exploiting the information given 
by $B_{\rm tag}$ allows for background 
suppression without assumptions on the 
(unknown)
signal dynamics.
We require zero total event charge as well as a negatively charged lepton with a momentum above 0.5 GeV/$c$ on the tagging side.
This reduces the main background,
where a $D_s^+$ produced in a decay of the type 
$B\to D^{(*)+}_s\bar{D}^{(*)}$
is combined with a lepton and a kaon from the subsequent $D$ decay in a semileptonic decay
$\bar{B}\to \ell^- {\bar \nu_\ell} D ^{(*)} X$ of 
the accompanying $\bar{B}$ meson.
Further improvement of the sensitivity is achieved with two tagging 
side variables $M^c_{\rm tag}\equiv \sqrt{(E_{\rm tag}-E^{~\ell}_{\rm tag})^2 - (\vec
p_{\rm tag}- \vec p_{\rm tag}^{~\ell})^2}$
and
$X_{\rm tag} \equiv  (E_{\rm beam} - E_{\rm tag} -
|\vec{p}_{\rm tag}|)/\sqrt{E^2_{\rm beam}-m_{B^+}^2}$,
where $E_{\rm tag}$ and $\vec{p}_{\rm tag}$ denote the 
total energy and momentum of all reconstructed particles not assigned to 
$B_{\rm sig}$, and $E_{\rm tag}^{~\ell}$ and $\vec{p}_{\rm tag}^{~\ell}$ 
represent the energy and momentum of the prompt tagging lepton.
Here $M^c_{\rm tag}$ represents the inclusively reconstructed mass
of the hadronic system produced in the $B_{\rm tag}$ decay 
and
$X_{\rm tag}$ is the tagging side equivalent of $X_{\rm mis}$.
The $M^c_{\rm tag}$ and $X_{\rm tag}$ 
distributions for signal and background are shown
in Fig.~\ref{tcuts}.

In this blind analysis, the selection criteria for $X_{\rm tag}$ and $M^c_{\rm tag}$ are optimized for the $D_s$ mode
by maximizing the expected statistical significance, ${N_S}/{\sqrt{N_S+N_B}}$, where
$N_S$ ($N_B$) is the predicted number of signal (background)
events in the 
($X_{\rm mis}$, $M_{D_s}$)  signal window.
This optimization is carried out 
for signal branching fractions
$\mathcal{B}(B^-\to D^+_s
K^-\ell^-\bar{\nu}_{\ell})$ in the range of
$(0.25 - 0.50) \times 10^{-3}$ and yields similar optimal selection criteria
for the whole range{, namely $-2<X_{\rm tag}<3$ and $M^c_{\rm tag}<2.4~{\rm GeV}/c^2$}.
$N_B$ is evaluated 
considering two background categories in the $D_s$ sample: 
``true $D_s$'' background
with correctly reconstructed $D_s^+$,
described by the 
MC scaled to the {integrated} luminosity in data, and a
``fake $D_s$'' component, where 
random track combinations are  misreconstructed as $D_s^+$, 
which 
is
evaluated from the $M_{D_s}$ sidebands.
In the $D^{*}_s$ sample, the background with true $D_s$ is split 
into two
parts:
``true $D^{*}_s$'' with properly reconstructed $D^{*+}_s$ and
``fake $D_s^{*}$'', where a true $D_s^+$ is combined with a 
random 
photon {candidate}. 
The background model 
is tested {using distributions}
in the sideband regions $X_{\rm mis}<-1$ and $X_{\rm mis}>1$. 

\begin{figure*} 
\includegraphics[width=0.2455\textwidth]{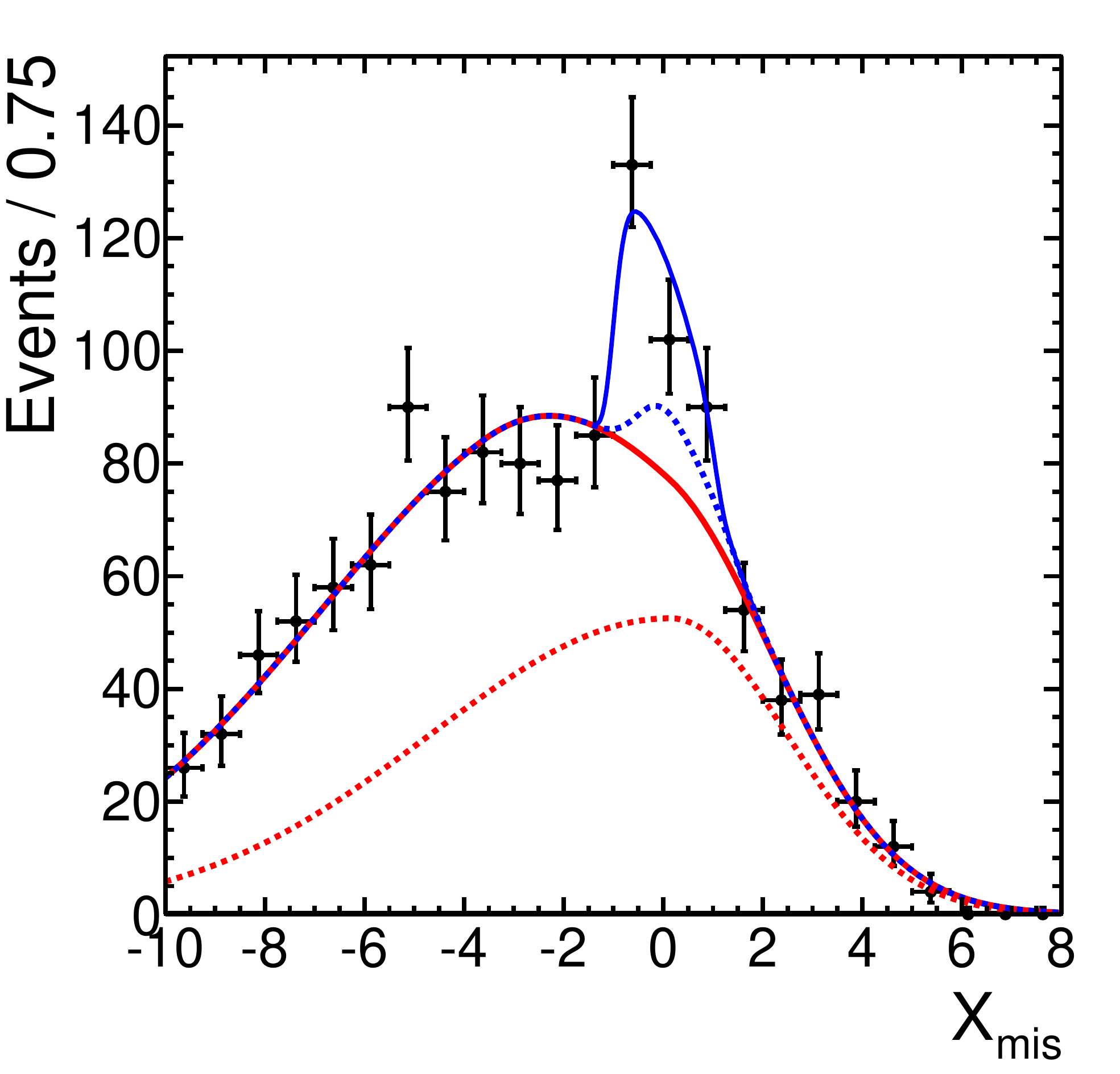}
\includegraphics[width=0.2455\textwidth]{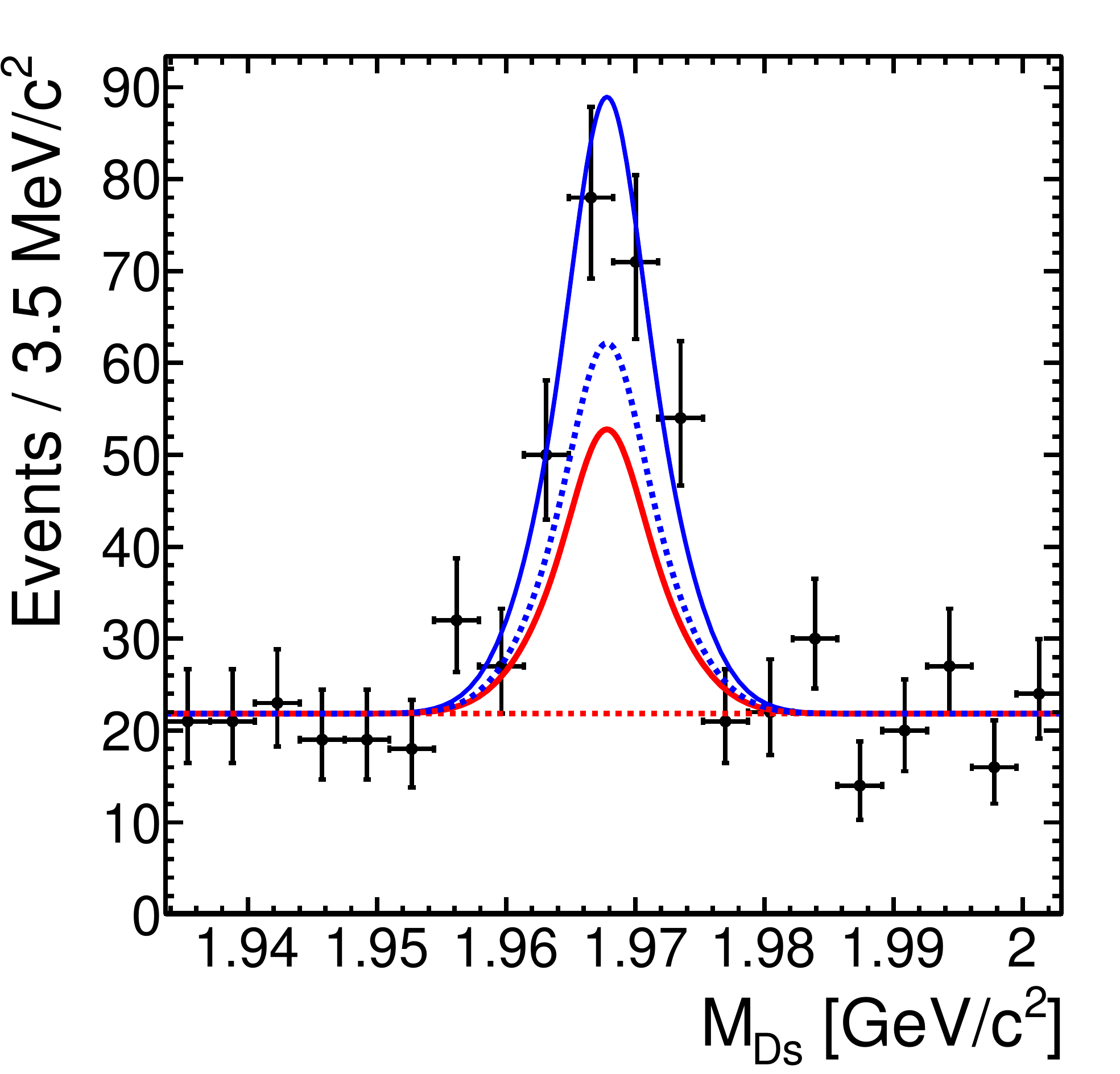} \\
\includegraphics[width=0.2455\textwidth]{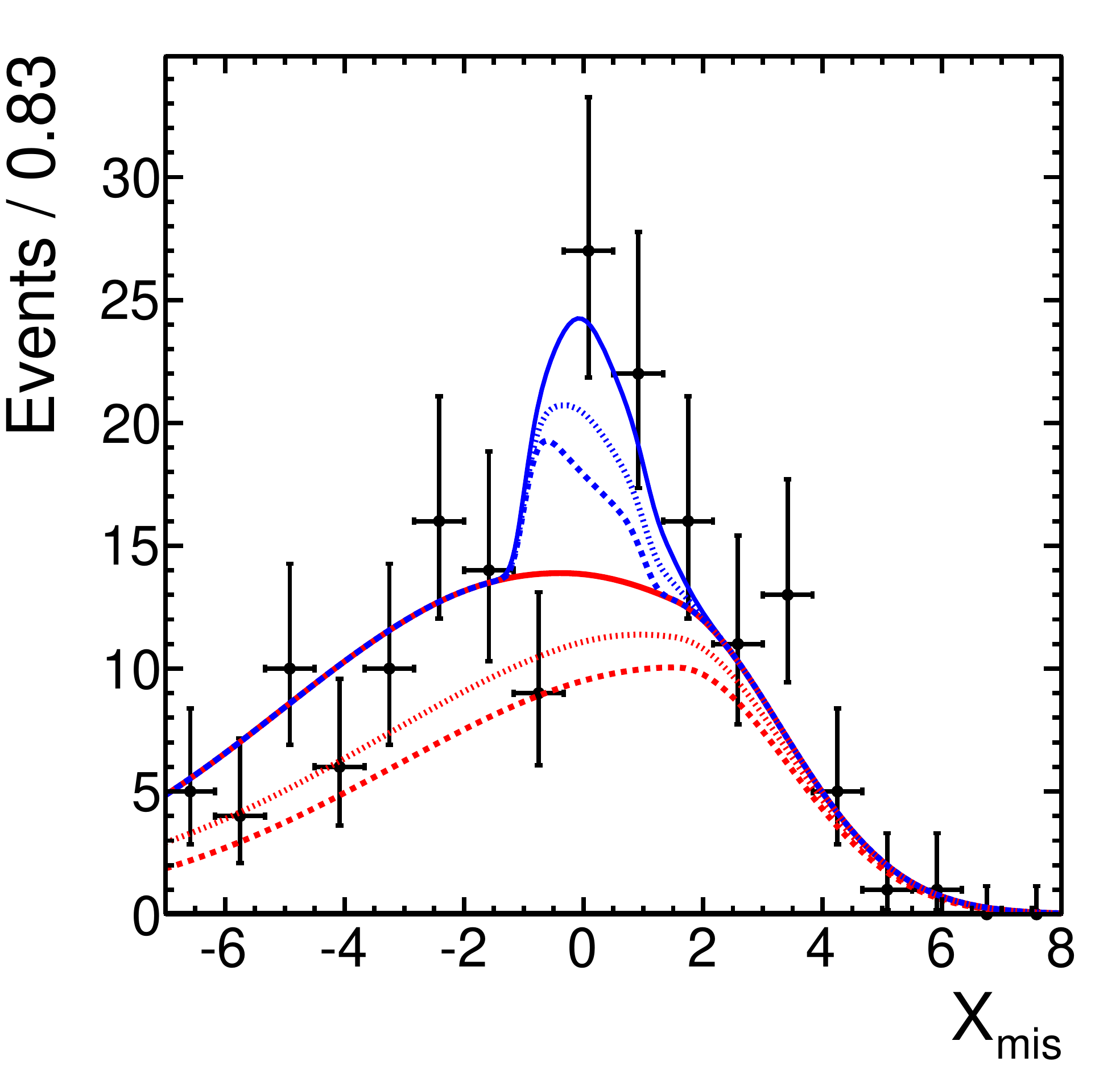}
\includegraphics[width=0.2455\textwidth]{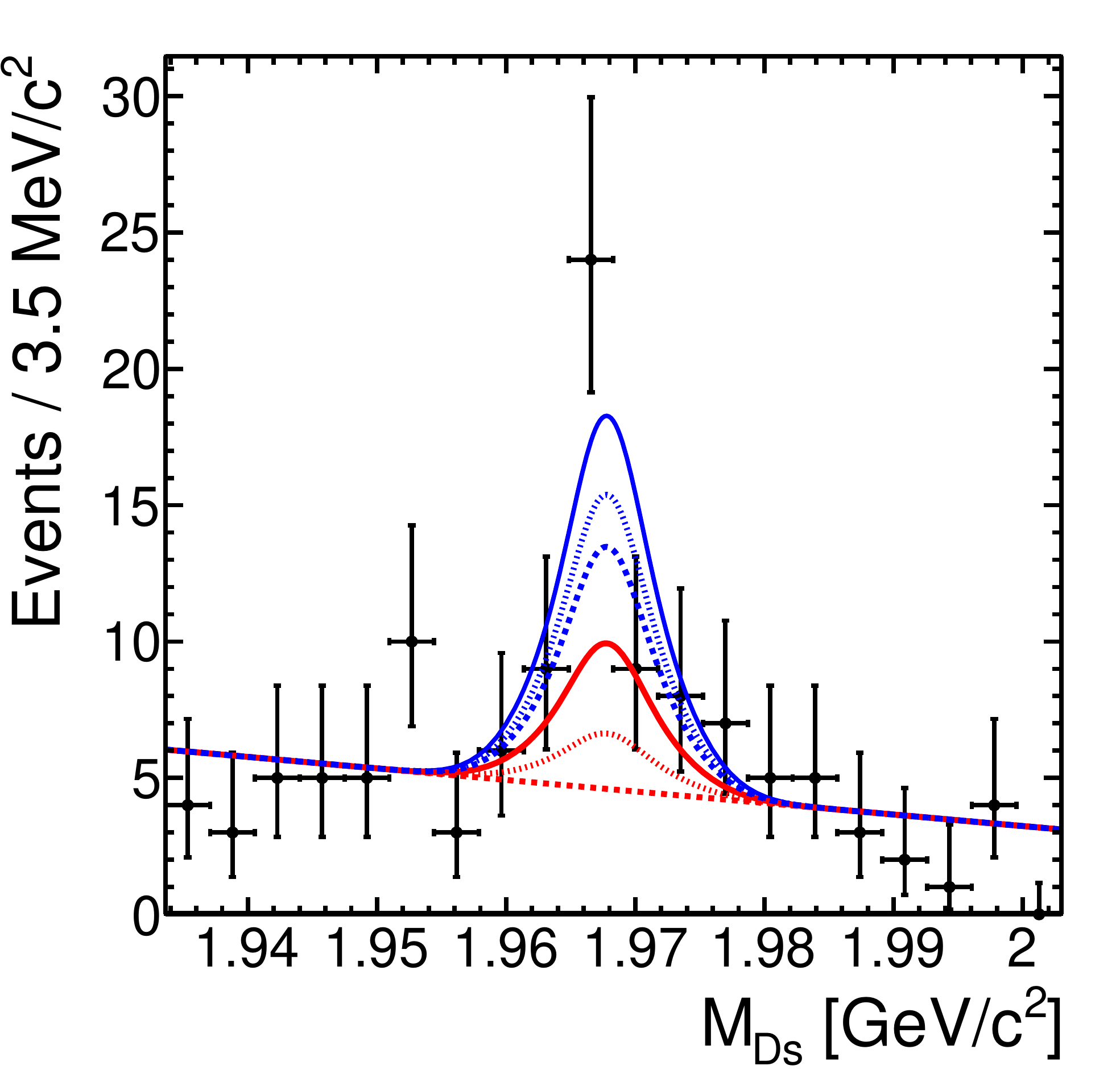}                
\includegraphics[width=0.2455\textwidth]{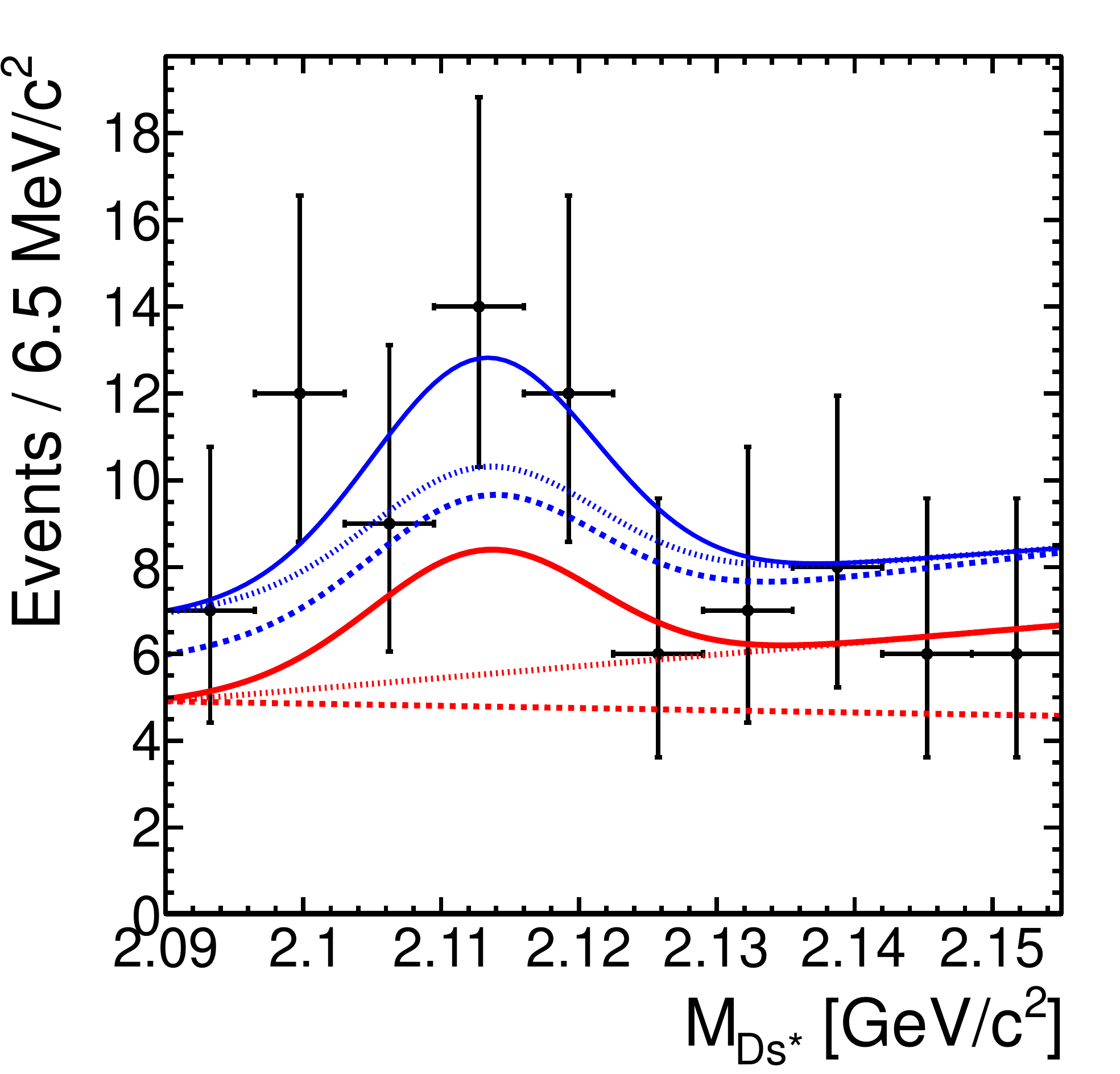}

\caption{Distributions (from left to right) in $X_{\rm mis}$
and $M_{D_s}$ in the $D_s$ sample {(top)}, and 
$X_{\rm mis}$, $M_{D_s}$ and $M_{D_s^*}$ in the $D_s^*$ sample (bottom).
Points with error bars are the data, and lines show the fit projections. Each variable is shown in the signal region of the other variable(s).
For the $D_s$ sample the lines represent (from bottom to top) 
the fitted background
components with fake (red dashed) and true $D_s$ (red solid),
and the signal contributions from the $D_s^{*}$ (blue dashed) and $D_s$ (blue solid) modes.
For the $D_s^{*}$ sample 
the lines (from bottom to top) represent the fitted
background components with fake $D_s$ (red dashed), fake $D_s^{*}$ (red dotted),
true $D_s^{*}$ (red solid), and the signal contributions
from the $D_s$ mode (blue dashed), the $D_s^{*}$ mode with fake $D_s^{*}$ (blue dotted),
and with true $D_s^{*}$ (blue solid).
The fitted contributions are superimposed additively.}
\label{all-fit}
\end{figure*}

\begin{figure} 
  \includegraphics[width=0.23\textwidth]{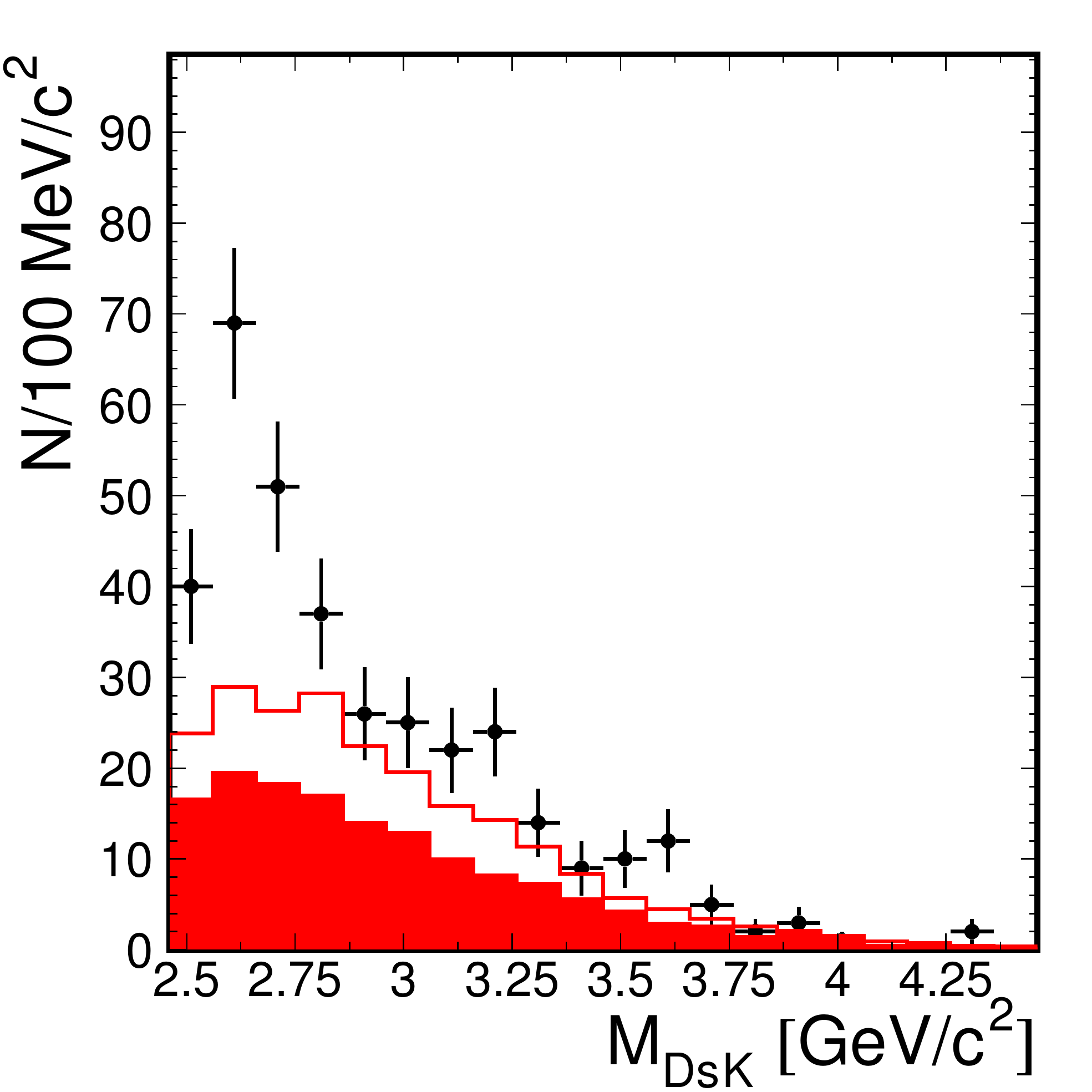}
\includegraphics[width=0.23\textwidth]{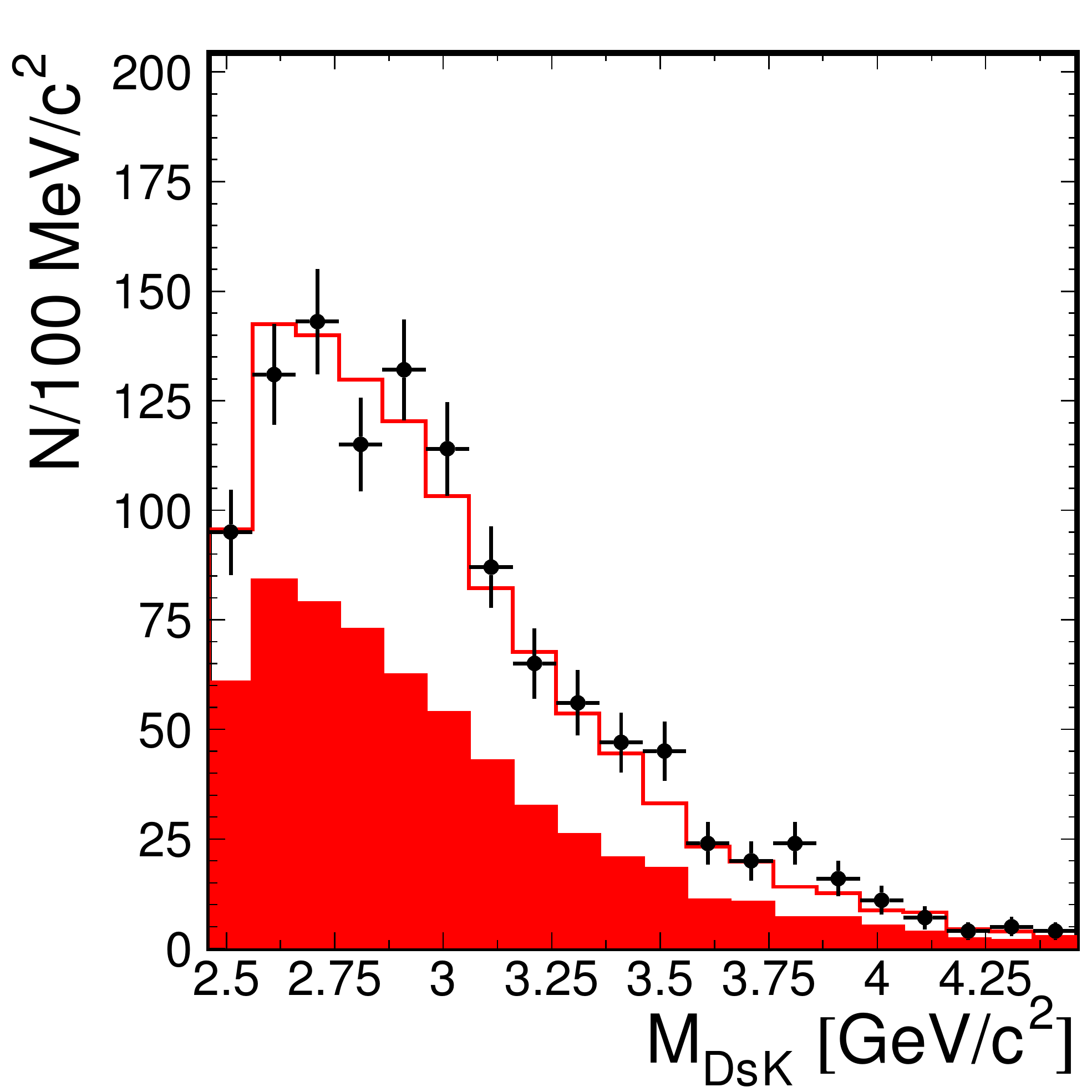}
\caption{The invariant mass distribution of $D_s^+K^-$ for the combined 
$D_s$ and $D_s^{*}$ samples
in the
signal window (left) and in the $X_{\rm mis}$ sidebands (right).
The full (blank) histograms show the expected background contribution from
fake (true) $D_s$. The histograms are superimposed additively.} 
\label{mdsk}
\end{figure}

The $X_{\rm mis}$ 
and $M_{D_s^{(*)}}$ 
distributions {in data} are shown in Fig.~\ref{all-fit}.
Figure~\ref{mdsk} shows the invariant mass distribution of the $D_s^+K^-$ 
system, $M_{D_sK}$, for the combined $D_s$ and $D_s^{*}$ samples 
in the signal window and in the $X_{\rm mis}$ sidebands.
Superimposed histograms represent the expected background{s}.
While the background model describes the experimental 
$M_{D_sK}$ distribution well
in the $X_{\rm mis}$ sidebands, 
a clear excess over the expected background 
is seen in the signal region.
The $M_{D_sK}$ distribution in the signal window is dominated
by a {prominent} peak
at $\approx 2.6$ GeV/$c^2$, similarly to that
observed in $B^-\to D_s^+K^-\pi^-$ decays
\cite{Wiechczynski}.

The signal yields are extracted from a simultaneous, extended unbinned 
maximum likelihood 
fit to the $D_s$ and $D_s^{*}$ samples{, consisting of 2175 and 396 events, respectively}.
The $D_s$ and $D^{*}_s$ samples are fitted 
in two ($X_{\rm mis}, M_{D_s}$) and three ($X_{\rm 
mis}, M_{D_s}, M_{D^{*}_s}$) dimensions, respectively. 
The 
likelihood function is 
constructed as follows: 
\begin{displaymath} \begin{array}{ll} \mathcal{L} 
= & e^{-(\sum_{k}N_k+\sum_{k'} N^{*}_{k'})} \prod^N_{i=1} [\sum_k N_k 
\mathcal{P}_k(x_i,y_i)]\times \\ & \prod^{N^{*}}_{i'=1} [\sum_{k'} 
N_{k'}^{*}\mathcal{P}_{k'}^{*}(x_{i'},y_{i'},z_{i'})] 
, 
\end{array}
\end{displaymath} 
where 
$x_{l}$, $y_{l}$, $z_l$ denote 
$X_{\rm mis}$, $M_{D_s}$ 
and 
$M_{D^{*}_s}$ in the $l^{\rm th}$ event, and $N^{(*)}$ denotes the 
total 
number 
of events in the $D^{(*)}_s$ data sample. The index $k$ ($k'$) runs 
over the signal and background components in the $D_s$ ($D_s^{*}$) 
sample; 
$N_{k}^{(*)}$ 
and $\mathcal{P}_{k}^{(*)}$ denote the number of events 
and the probability density functions (PDF) for each component, respectively. In the 
$D_s$ sample, we consider two signal components coming  
from the {decay $B^-\to D^+_sK^-\ell^-\bar{\nu}_{\ell}$ and from the decay $B^-\to 
D^{*+}_sK^-\ell^-\bar{\nu}_{\ell}$ if a photon from the $D^{*+}_s$
has been missed.}
In the $D_s^{*}$ sample, we distinguish three signal components: one 
coming from the $B^-\to D^+_sK^-\ell^-\bar{\nu}_{\ell}$ mode, where 
 the $D_s$ 
meson is associated with a random photon, and two from the $B^-\to 
D^{*+}_sK^-\ell^-\bar{\nu}_{\ell}$ mode, with true and fake 
$D_s^{*}$ defined  similarly to the background case  
discussed above. 
The coefficients 
$N_{k}^{(*)}$ for the signal components are expressed as the products 
$N_{k}^{(*)}=N_{D_s^{(*)}}f^{(*)}_{k}$, where 
$N_{D_s^{(*)}}$ denotes the 
total number of signal events in the $B^-\to D^{(*)+}_s 
K^-\ell^-\bar{\nu}_{\ell}$ modes.  The coefficients $f^{(*)}_{k}$ {(listed in Table \ref{fk})}
represent
the signal fraction reconstructed in each component and are evaluated from 
the signal MC. 
\begin{table}
\caption{The coefficients $f^{(*)}_{k}$, representing the signal fraction
reconstructed in each component, evaluated from the signal MC.
}
\begin{tabular}{l c c}
\hline\hline
Signal component $k$ & Sample & $f^{(*)}_{k}$ \\
\hline
\multirow{2}{*}{$B^-\to D_s^{+}K^-\ell^-\bar{\nu}_{\ell}$} & $D_s$ & $(84 \pm 1)\%$ \\
	& $D_s^*$ & $(16 \pm 1)\%$ \\
	\hline
\multirow{3}{*}{$B^-\to D_s^{*+}K^-\ell^-\bar{\nu}_{\ell}$} & $D_s^*$ with true $D_s^*$ & $(21 \pm 1)\%$ \\
	& $D_s^*$ with fake $D_s^*$ & $(13 \pm 1)\%$ \\
	& $D_s$ & $(66 \pm 1)\%$ \\
\hline\hline
\end{tabular}
\label{fk}
\end{table}
The 
coefficients $N_{k}^{(*)}$ for background components with fake $D_s$ 
are 
evaluated from the $M_{D_s}$ sidebands in data and are fixed in the fit.  The 
two- (three-) dimensional PDF is parameterized as the 
product of 
two (three) one-dimensional PDFs for each variable. 
The validity of this
parameterization has been checked with MC by examining the
{correlation between $X_{\rm mis}$ and $M_{D_s}$, which has been found negligible.}
The components with true 
$D_s^{(*)}$ are parameterized as a sum of two Gaussian 
{functions} in $M_{D_s}$ or as 
a single Gaussian 
{function} 
in $M_{D_s^{*}}$, with means set to the world average 
$D^{(*)}_s$ mass values \cite{PDG} and with the
remaining parameters 
fixed from fits to 
control samples in data. The components with fake $D_s^{(*)}$ are 
parameterized as linear functions in $M_{D^{(*)}_s}$. 
The $X_{\rm mis}$ {distribution} of the signal components is modeled with 
two line shapes, one describing the two components of the $B^-\to 
D^{+}_sK^-\ell^-\bar{\nu}_{\ell}$ 
mode and the other one describing the three components of the $B^-\to 
D^{*+}_sK^-\ell^-\bar{\nu}_{\ell}$ 
decay. 
They are parameterized using the function
$Ce^{-|(X_{\rm mis}-\mu)/{\sigma}|^n} e^{-\alpha (X_{\rm mis} -\mu)}$,
where $C$ is a normalization 
coefficient and the parameters $\mu$, $\sigma$, 
$\alpha$ and the integer parameter $n$ 
are fixed from fits to the signal MC samples.
The $X_{\rm mis}$ distributions of the background components are 
parameterized as bifurcated Gaussian functions  with 
parameters fixed from 
the simulated $B\bar{B}$ events with generic $B$ decays 
(true $D_s$) or from the $M_{D_s}$ 
sidebands in data (fake 
$D_s$).
The free parameters in the fit are the two signal 
yields $N_{D^{(*)}_s}$, the three  
background yields $N^{(*)}_m$ of the  
components with true $D_s$, and the coefficients of polynomials that describe the distributions in
$M_{D_s^{(*)}}$ for the fake $D_s$ components. 
The range of the fit is as shown in Fig.~\ref{all-fit}.
The signal yields extracted from the fit are $84\pm24$ events 
for the
decay $B^-\to D^+_sK^-\ell^-\bar{\nu}_{\ell}$ and 
$41\pm22$ events 
for the decay $B^-\to D^{*+}_sK^-\ell^-\bar{\nu}_{\ell}$
with statistical significances of 3.9{$\sigma$} and 1.9{$\sigma$}, 
respectively. 
{The s}ignificance is defined as $\Sigma = 
\sqrt{-2{\ln}(\mathcal{L}_{\rm 0}/\mathcal{L}_{\rm max})}$, 
where $\mathcal{L}_{\rm max}$ 
and $\mathcal{L}_{\rm 0}$ denote the maximum likelihood value 
and the likelihood value for the zero signal hypothesis, respectively.
The fit results are summarized in Table \ref{fit-results} and
the fit projections in $X_{\rm mis}$ {and $M_{D_s}$}
are shown 
in Fig.~\ref{all-fit}.
\begin{table}
\caption{Signal yields ($N_{D_s^{(*)}}$), reconstruction 
efficiencies ($\epsilon^{(*)}$), statistical  
significances ($\Sigma$) and branching fractions ($\mathcal{B}$).
The errors on the signal yields are statistical, while for the branching fractions both statistical (first) and systematic (second) errors are provided. The correlation coefficient between $N_{D_s}$ and $N_{D_s^{*}}$ equals $-66\%$.
}
\begin{tabular}{l c c c c}
\hline\hline
Mode & $N_{D_s^{(*)}}$ & $\epsilon^{(*)} [\%]$&$\Sigma$&
$\mathcal{B}\times 10^{-3}$\\
\hline
$B^-\to D_s^+K^-\ell^-\bar{\nu}_{\ell}$ &$84\pm 24$&1.78&3.9&
$0.30\pm 0.09^{+0.11}_{-0.08}$\\
 
$B^-\to D_s^{*+}K^-\ell^-\bar{\nu}_{\ell}$&$41\pm 22$&0.85&1.9
& $0.29 \pm 0.16^{+0.11}_{-0.10}$\\
\hline\hline
\end{tabular}
\label{fit-results}
\end{table}
{The fitted signal yields are used to compute the branching fractions with the formula:}
$\mathcal{B}(B^-\to D_s^{(*)+}K^-\ell^-\bar{\nu}_{\ell}) = 
{N_s^{(*)}}/({2N_{B^+B^-} \epsilon^{(*)}\mathcal{B}_{int}})$,
where $N_{B^+B^-}$ is the number of 
$B^+B^-$ pairs in data, $\epsilon^{(*)}$ 
denotes the reconstruction 
efficiency of the signal decay chain and $\mathcal{B}_{int}$ is 
the product of intermediate branching 
fractions set to their world average values \cite{PDG}.
The reconstruction efficiency is expressed as
$\epsilon^{(*)} = \epsilon_{PS}^{(*)}\Delta\epsilon_{cor}^{(*)}$,
where $\epsilon_{PS}^{(*)}$ is the efficiency calculated from the signal MC with 
the phase space model and
$\Delta\epsilon_{cor}^{(*)} = 1.20$ ($0.57$) corrects for the difference 
between the data 
and the phase space distribution.
It is calculated as a 
function of {the}
effective 
masses of {the} two-body subsystems $D^+_sK^-$,
$D^+_s\ell^-$, and $K^-\ell^-$
and averaged using the 
experimentally observed distributions. 
We obtain
$\mathcal{B}(B^-\to D^{+}_sK^-\ell^-\bar{\nu}_{\ell} ) 
= (0.30\pm0.09)\times 10^{-3}$ and 
$\mathcal{B}(B^-\to D^{*+}_s K^-\ell^-\bar{\nu}_{\ell})
=(0.29\pm 0.16)\times 10^{-3}$.

The dominant systematic uncertainty on the signal yield is due to the parameterization of the $X_{\rm mis}$ dependence of the signal and found to be $^{+23}_{-~6}$($^{+7}_{-9}$) events for the $D_s$($D_s^{*}$) mode.
It is evaluated by refitting the data 
with the parameters $\mu$, $\sigma$, and $\alpha$ allowed 
to float, and by changing the integer parameter $n$ by $\pm 1$.
Uncertainties in modeling the $X_{\rm mis}$ distributions of the
background components containing true $D_s$ are evaluated to be 
$^{+5}_{-7}$ ($^{+8}_{-7}$) events
from fits  
with the background shape parameters varied by  
$\pm 1\sigma$, taking into account correlations between the parameters. 
We also repeat the fits with the parameters, 
whose values are determined from data (and which are fixed in the 
nominal fit), floating.
The resulting uncertainty is $^{+4}_{-2}$ ($_{-1}^{+0}$) events.
The effect of an imperfect estimation of the relative contributions of 
the signal components 
is determined to be $\pm 1$ ($\pm 1$) from fits with the parameters 
$f_k^{(*)}$ varied by $\pm 1\sigma$ and  
taking into account a $\pm3\%$ uncertainty on the photon reconstruction efficiency.
The above uncertainties 
are summed in quadrature to obtain 
the total systematic uncertainty of the signal yield
of $^{+24}_{-10}$ ($^{+12}_{-11})$ events for the $D_s$ ($D_s^{*}$) modes.
We include the effect of these uncertainties 
on the significance of the observed signals by convolving the 
likelihood function obtained in the fit  with 
a Gaussian systematic error 
distribution. The significance of the signal in the 
$B^-\to D^+_sK^-\ell^-\bar{\nu}_{\ell}$ ($B^-\to 
D^{*+}_sK^-\ell^-\bar{\nu}_{\ell}$) 
mode, 
after including 
systematic uncertainties, is 3.4$\sigma$ (1.8$\sigma$).

In a similar way, we obtain a significance of 6$\sigma$ for 
the combined $B^-\to D^{(*)+}_sK^-\ell^-\bar{\nu}_{\ell}$ 
modes from the 2-dimensional ($X_{\rm mis}$, $M_{D_s}$) 
fit for the combined $D_s$ and $D_s^{*}$ samples.
The much higher significance for the combined modes compared to the individual modes
is due to the large cross-feed between the $D_s$ and the $D_s^*$ modes.

  The uncertainty on
the branching fractions, except for
the systematic uncertainty of the signal yield,
is evaluated to be $23.2\%$ for each signal mode.
It includes uncertainties in charged track reconstruction efficiency ($6.6\%$),
particle identification efficiency ($3.9\%$), intermediate branching fractions 
($6.1\%)$,
number of $B^+B^-$ pairs ($1.5\%$) and 
the reconstruction efficiency correction $\Delta\epsilon_{cor}$ ($21\%$).

The largest uncertainty, due to $\Delta\epsilon_{cor}$, is determined by calculating $\Delta\epsilon_{cor}$ in 10000 toy MC experiments.
The width of a Gaussian function fitted to 
the obtained efficiencies is {taken} as systematic uncertainty.   
The uncertainties due to the intermediate branching 
fractions 
are taken from the errors quoted in \cite{PDG}.  
Combining all uncertainties, we obtain 
$\mathcal{B}(B^-\to D^+_sK^-\ell^-\bar{\nu}_{\ell})=
(0.30\pm 0.09({\rm stat})^{+0.11}_{-0.08}({\rm syst}))\times 10^{-3}$,
$\mathcal{B}(B^-\to D^{*+}_sK^-\ell^-\bar{\nu}_{\ell})=
(0.29\pm 0.16({\rm stat})^{+0.11}_{-0.10}({\rm syst}))\times 10^{-3}$ and $\mathcal{B}(B^-\to D^{(*)+}_sK^-\ell^-\bar{\nu}_{\ell})=
(0.59\pm 0.12({\rm stat})\pm 0.15({\rm syst}))\times 10^{-3}$ for the combined modes obtained in a similar way, taking correlations into account.
Since the significance in the
$D^{*}_s$ mode
does not exceed $3\sigma$, we 
set an upper limit of
$\mathcal{B}(B^-\to D^{*+}_sK^-\ell^-\bar{\nu}_{\ell})<0.56\times 
10^{-3}$
at the $90\%$ confidence level, using the likelihood integration method.

In conclusion, we find evidence for the decay
$B^-\to D^{+}_sK^-\ell^-\bar{\nu}_{\ell}$ with a significance of 
$3.4\sigma$
and measure  
$\mathcal{B}(B^-\to D^+_sK^-\ell^-\bar{\nu}_{\ell})=
(0.30\pm 0.09({\rm stat})^{+0.11}_{-0.08}({\rm syst}))\times 10^{-3}$.  The combined $B^-\to D^{(*)+}_sK^-\ell^-\bar{\nu}_{\ell}$  
decay modes are observed with a significance of $6\sigma$ to be $\mathcal{B}(B^-\to D^{(*)+}_sK^-\ell^-\bar{\nu}_{\ell})=
(0.59\pm 0.12({\rm stat})\pm 0.15({\rm syst}))\times 10^{-3}$.
The 
branching fraction results
are consistent with the 
measurement of BaBar \cite{BaBar}.
We also present the first measurement of the $D^+_sK^-$ 
invariant mass distribution,
which is dominated by a prominent {peak
around} $2.6$ GeV$/c^2$, possibly from excited $D$ mesons decays. 

We thank the KEKB group for excellent operation of the
accelerator; the KEK cryogenics group for efficient solenoid
operations; and the KEK computer group, the NII, and 
PNNL/EMSL for valuable computing and SINET4 network support.  
We acknowledge support from MEXT, JSPS and Nagoya's TLPRC (Japan);
ARC and DIISR (Australia); NSFC (China); MSMT (Czechia);
DST (India); INFN (Italy); MEST, NRF, GSDC of KISTI, and WCU (Korea); 
MNiSW and NCN (Poland); MES and RFAAE (Russia); ARRS (Slovenia); 
SNSF (Switzerland); NSC and MOE (Taiwan); and DOE and NSF (USA).


\end{document}